\documentclass[showpacs,amssymb,aps,twocolumn]{revtex4}
\usepackage{amsmath}
\usepackage{graphicx}
\usepackage{epsf}
\usepackage{graphics}
\usepackage[latin1]{inputenc}
\newcommand{\be}{\begin{equation}}
\newcommand{\ee}{\end{equation}}
\newcommand{\bea}{\begin{eqnarray}}
\newcommand{\eea}{\end{eqnarray}}

\begin{document}

\title{Instabilities in Thermal Gravity with a Cosmological Constant}

\author{F. T. Brandt and  J. Frenkel}
\address{Instituto de Física, Universidade de São Paulo, 05508-090, São Paulo, SP, BRAZIL}

\begin{abstract}
It is shown that in quantum gravity at finite temperature, the effective potential
evaluated in the tadpole approximation can have a local minimum 
below a certain critical temperature. However, when the leading
higher order thermal loop corrections are included, one finds that no
static solution exists at high temperature.
\end{abstract}


\maketitle

\section{Introduction}

The purely attractive nature of gravity is the source of many
instabilities, both at the classical and quantum levels. Instabilities
in quantum gravity at finite temperature have been studied from various points of 
view \cite{Gross:1982cv,kikuchi:1984np,Nakazawa:1984zq,gribosky:1989yk,Frenkel:1991dw,Rebhan:1991yr,Brandt:1992qn,Barrow:2003ni}.
Here we will consider this aspect in the
spirit discussed by several authors \cite{LeBellac:1989ps,Grandou:1990ir,Pisarski:1990ds,Altherr:1991fu}
in the context of the $\left(\phi^3\right)_6$ scalar model.

It is well known that at zero temperature, a gravitational system which
would naturally fall upon itself, may remain in a static and uniform
state in the presence of a cosmological constant. It may therefore be
interesting to inquire whether such a situation can hold also at
non-zero temperature.
In this note, 
we consider the leading high temperature contributions to the
effective potential in quantum gravity, assuming that
self-interactions of the thermal matter are negligible. We show that,
in this regime, the effective potential has no static solution.
In the approximation where loops appear only
through tadpole (one-point) diagrams we find that, for negative values
of the cosmological constant $\Lambda$, the effective potential can
have a local minimum below a critical temperature $T_{\rm cr}$.
These features are  somewhat similar to the ones encountered
in the $\left(\phi^3\right)_6$ 
scalar model at finite temperature \cite{Altherr:1991fu}.
But, in contrast with this theory, quantum gravity has a gauge invariance
which requires that loops with any number of external lines must have
the same $T^4$ leading dependence at high temperature
\cite{Frenkel:1991dw,Rebhan:1991yr,Brandt:1992qn}. When all such loop
contributions are taken into account, it turns out that the effective
potential exhibits instabilities at high temperatures.
The terms which destabilize the local minimum of the effective
potential are those associated with the hard thermal self-energy and
vertex corrections to the tree amplitudes which join together the
tadpole graphs. Such an instability indicates that, even in the
presence of a cosmological constant, 
it is likely that a stable gravitational system cannot be formed at
high temperature.

In section 2 we study the effective potential at finite temperature in
the tadpole approximation, while in section 3 we discuss the effect of
the inclusion of higher order thermal loop corrections. Some results
which follow from this analysis may also be understood by means of a
diagrammatic approach.

\section{Effective potential in the tadpole approximation}
The effective potential $V_{\rm eff}$, which is associated with configurations
involving constant fields, is a useful quantity for studying the
stability of a thermodynamic system
\cite{Coleman:1973jx,dolan:1974qd,weinberg:1974hy,Fujimoto:1984yz,evans:1987ws}.
In particular, the vacuum expectation values of the fields may 
be determined by minimizing $V_{\rm eff}$. Since the Ricci scalar vanishes for constant fields, we
can write the effective potential in the form
\be\label{eq01}
V_{\rm eff} = \frac{\Lambda}{8\pi G}\sqrt{-g} + {\cal{H}},
\ee
where ${\cal{H}}$ describes 
the matter fields coupled to gravity as well as
the hard thermal effects.

In this section we shall consider the effective potential in the
tadpole approximation, which is obtained by keeping in ${\cal{H}}$ only
the linear term in the gravitational field. Then, one may express the
effective potential as
\be\label{eq02}
V_{\rm eff} = \frac{\Lambda}{8\pi G}\sqrt{-g} + \frac{1}{2} T_0^{\mu\nu} g_{\mu\nu}
\ee
where $T_0^{\mu\nu}$ denotes the energy-momentum tensor of the system
which includes the leading thermal contributions to lowest
order. For example, in the presence of a constant scalar field
$\phi_c$ with mass $m$,
$T_0^{\mu\nu}$ can be written in the rest frame of the thermal bath as
\be\label{eq03}
T_0^{\mu\nu} = \frac{m^2 \phi_c^2}{2}\eta^{\mu\nu} +
\omega \frac{\pi^2 T^4}{90}\left(4\eta^{\mu 0} \eta^{\nu 0} - \eta^{\mu\nu} \right),
\ee
where $\omega$ is a numerical factor which depends on the quantum numbers of the thermal fields.

The above effective potential will have a local minimum at 
\be\label{eq04}
2 \frac{\delta V_{\rm eff}}{\delta g_{\mu\nu}} =
\frac{\Lambda}{8\pi G} \sqrt{-g} g^{\mu\nu} + T_0^{\mu\nu} =0. 
\ee
This equation can be solved in closed form and the solution, which
corresponds to the vacuum expectation value $\langle g^{\mu\nu}\rangle$,
is given by
\be\label{eq05}
\langle g^{\mu\nu}\rangle = -
\frac{\Lambda}{8\pi G} \frac{T_0^{\mu\nu}}{\sqrt{-\det T_0}} .
\ee
One can give a simple diagrammatic interpretation of this result, which
may illustrate some points. To this end, let us define
\be\label{eq06}
g_{\mu\nu} = \eta_{\mu\nu} + \kappa
h_{\mu\nu},\;\;\;\kappa\equiv\sqrt{32\pi G}
\ee 
and expand $V_{\rm eff}$ as a power series in the field $h_{\mu\nu}$. Then, up to
constants, we can write $V_{\rm eff}$ in the form
\begin{eqnarray}\label{eq07}
V_{\rm eff} & = & \frac{\kappa}{2} \bar T_0^{\mu\nu} h_{\mu\nu} 
+\frac{1}{2} h_{\mu\nu} D^{-1\;\mu\nu;\alpha\beta} h_{\alpha\beta}
\nonumber \\
&+&\frac{\kappa}{3!} \Gamma_3^{\mu\nu;\alpha\beta;\rho\sigma} h_{\mu\nu} h_{\alpha\beta} h_{\rho\sigma}  
\nonumber \\
&+&\frac{\kappa^2}{4!} \Gamma_4^{\mu\nu;\alpha\beta;\rho\sigma;\lambda\delta}  
h_{\mu\nu} h_{\alpha\beta} h_{\rho\sigma}  h_{\lambda\delta}  + \cdots .
\end{eqnarray}
Here $\bar T_0^{\mu\nu}$ denotes the effective energy-momentum tensor
\be\label{eq08}
\bar T_0^{\mu\nu} \equiv T_0^{\mu\nu} + \frac{\Lambda}{8\pi G}\eta^{\mu\nu}.
\ee
$D^{-1\;\mu\nu;\alpha\beta}$ is the inverse of the graviton propagator and $\Gamma_3$,
$\Gamma_4$, $\cdots$ represent the graviton interaction vertices at
zero momenta.
Then, the Euler-Lagrange equation \eqref{eq04} can be written as
\begin{eqnarray}\label{eq09}
&&{D}^{-1\mu\nu;\alpha\beta}h_{\alpha\beta}+
\frac{\kappa}{2!}\Gamma_3^{\mu\nu;\alpha\beta;\rho\sigma}h_{\alpha\beta} h_{\rho\sigma}+
\nonumber \\ &+&
\frac{\kappa^2}{3!}\Gamma_4^{\mu\nu;\alpha\beta;\rho\sigma;\lambda\delta}
h_{\alpha\beta} h_{\rho\sigma} h_{\lambda\delta}+\cdots = 
-\frac{1}{2}\kappa\bar T_0^{\mu\nu} .
\end{eqnarray}
Using the Dyson-Schwinger method, we can now obtain a perturbative
solution of \eqref{eq09} as a power series in $\bar T_0$, which may be
represented graphically as
\begin{eqnarray}\label{eq10}
\langle h_{\mu\nu} \rangle 
&=&
- \frac{1}{2} \begin{array}{c}\includegraphics[scale=0.2]{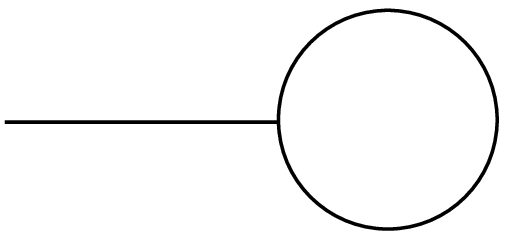}\end{array}
-\frac{1}{8} \begin{array}{c}\includegraphics[scale=0.2]{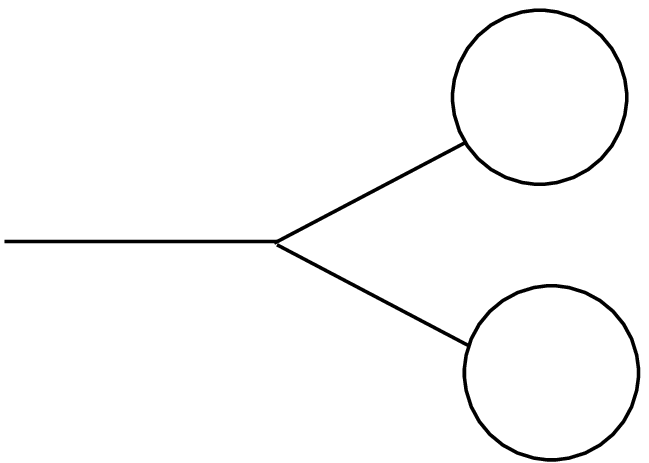}\end{array}
\nonumber \\
&-&\frac{1}{16} \begin{array}{c}\includegraphics[scale=0.2]{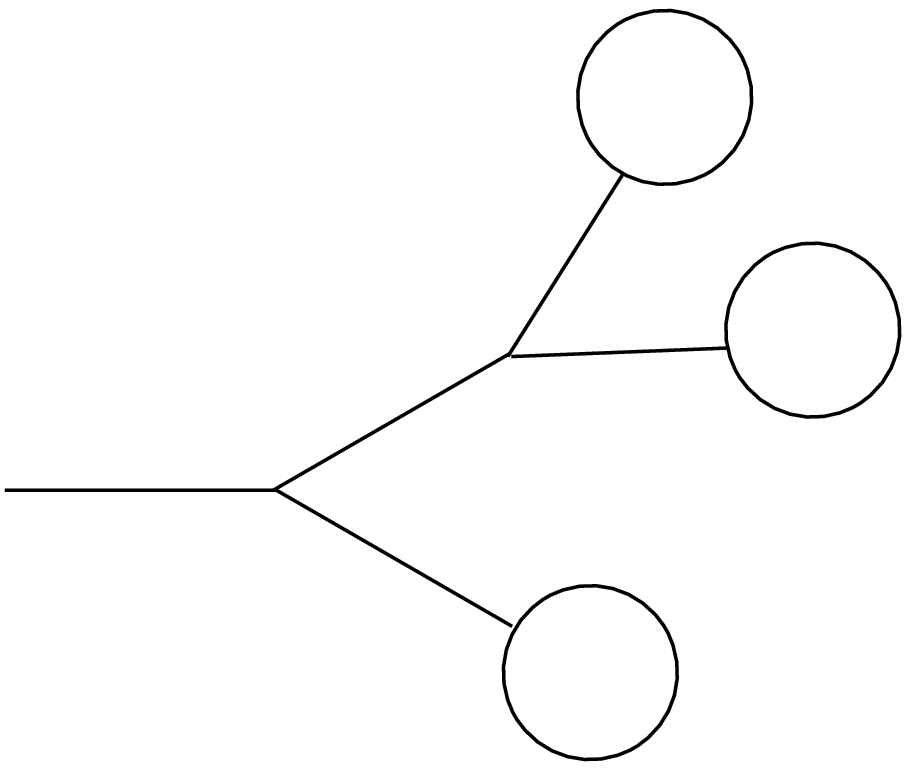}\end{array}
+\frac{1}{48} \begin{array}{c}\includegraphics[scale=0.2]{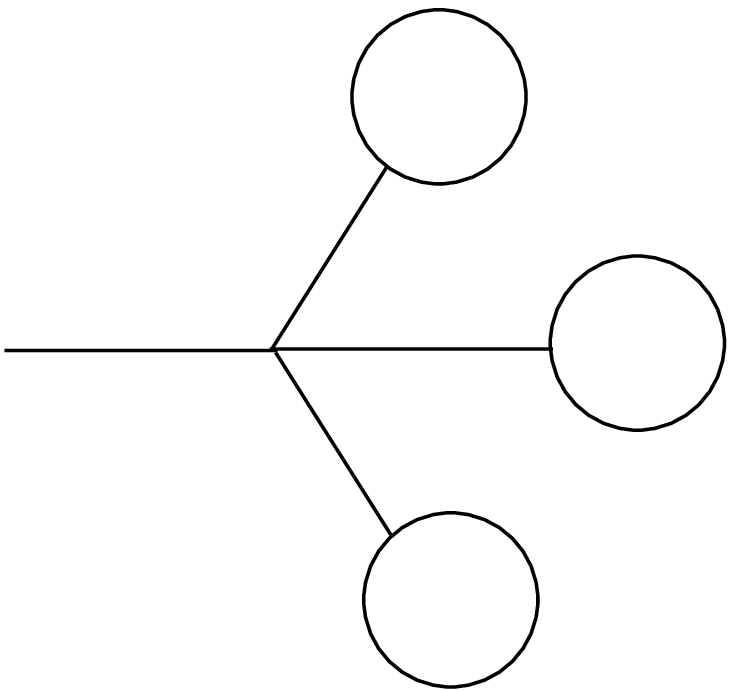}\end{array}
\nonumber \\
&-&\frac{1}{128} \begin{array}{c}\includegraphics[scale=0.2]{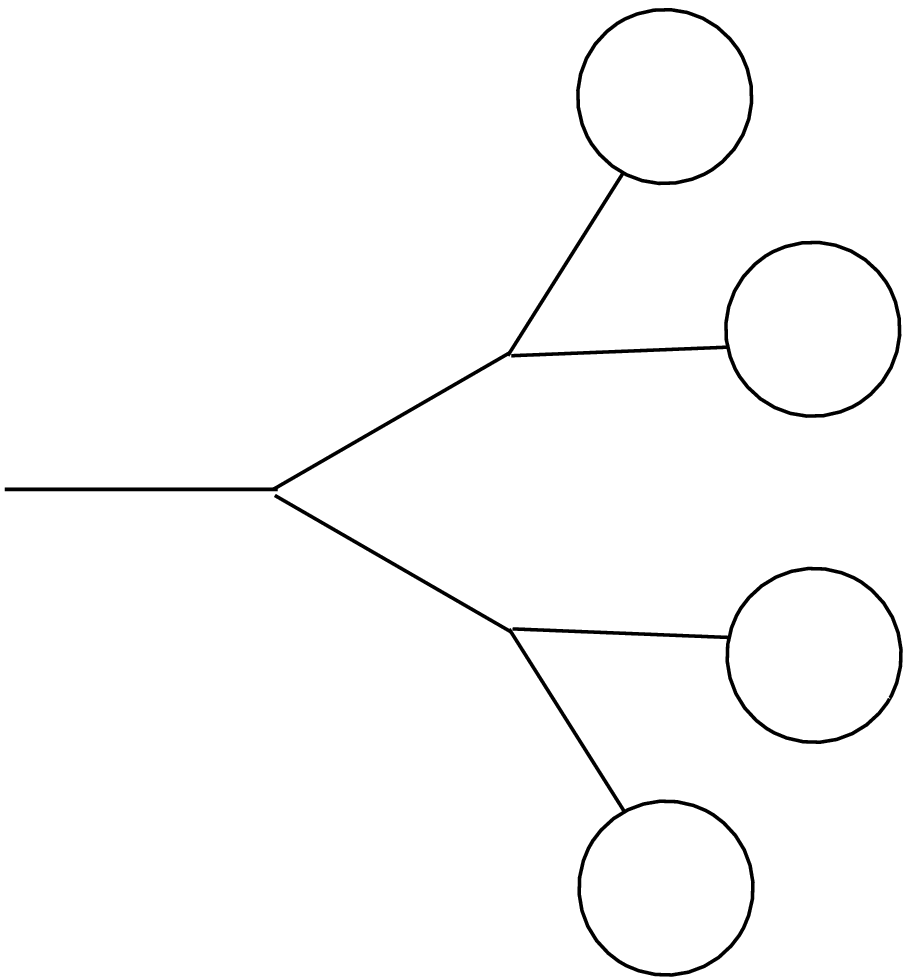}\end{array}
+\frac{1}{64} \begin{array}{c}\includegraphics[scale=0.2]{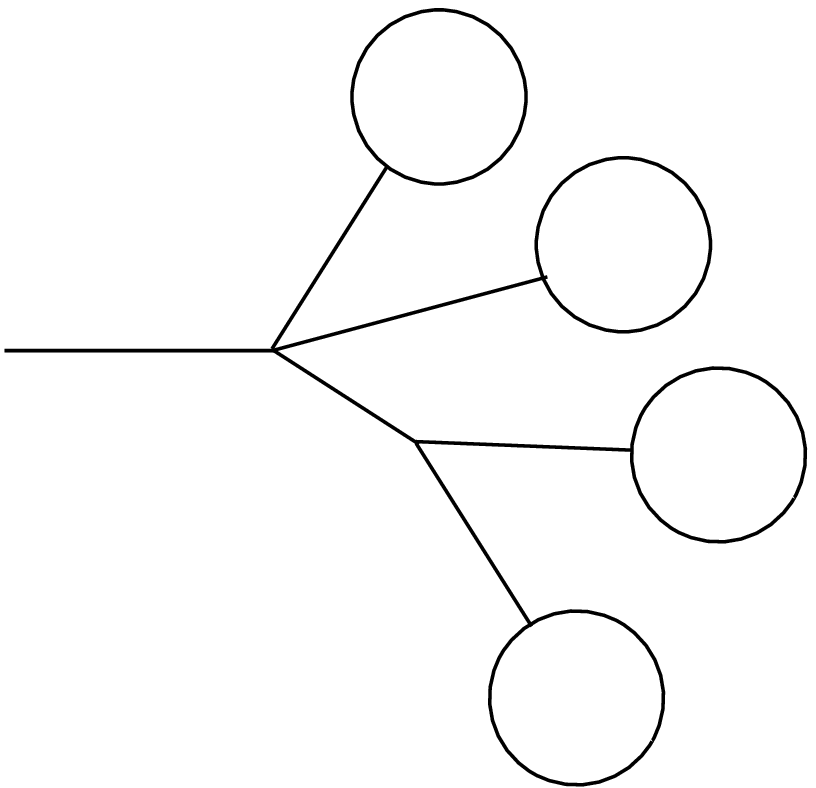}\end{array}
\nonumber \\
&-&\frac{1}{384}\begin{array}{c}\includegraphics[scale=0.2]{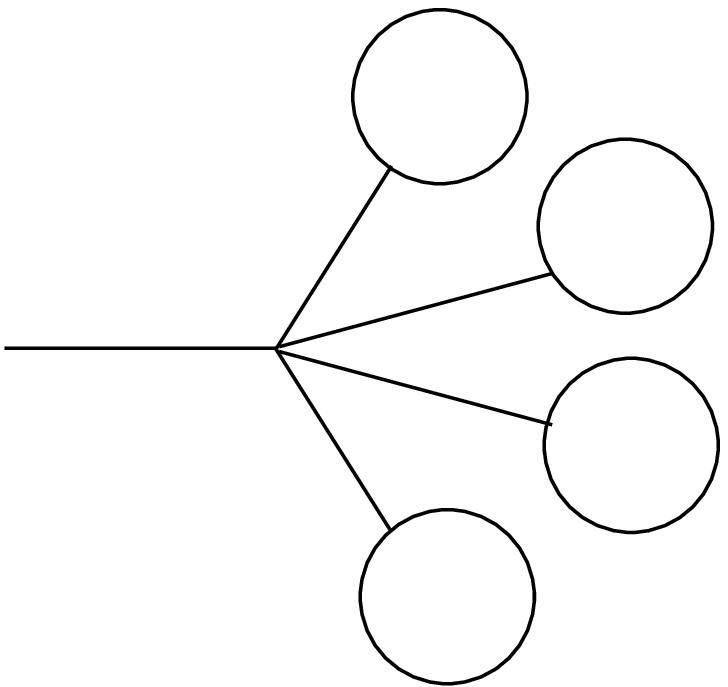}\end{array}
-\frac{1}{32} \begin{array}{c}\includegraphics[scale=0.2]{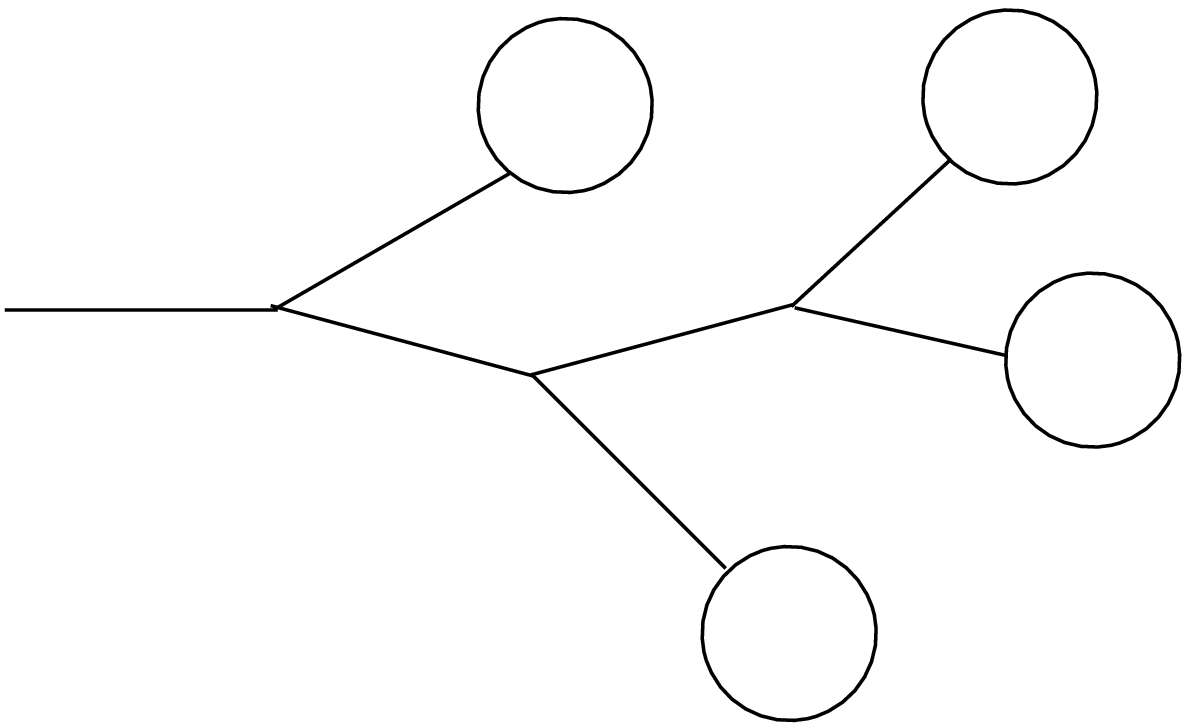}\end{array}
\nonumber \\
&+&\frac{1}{96} \begin{array}{c}\includegraphics[scale=0.2]{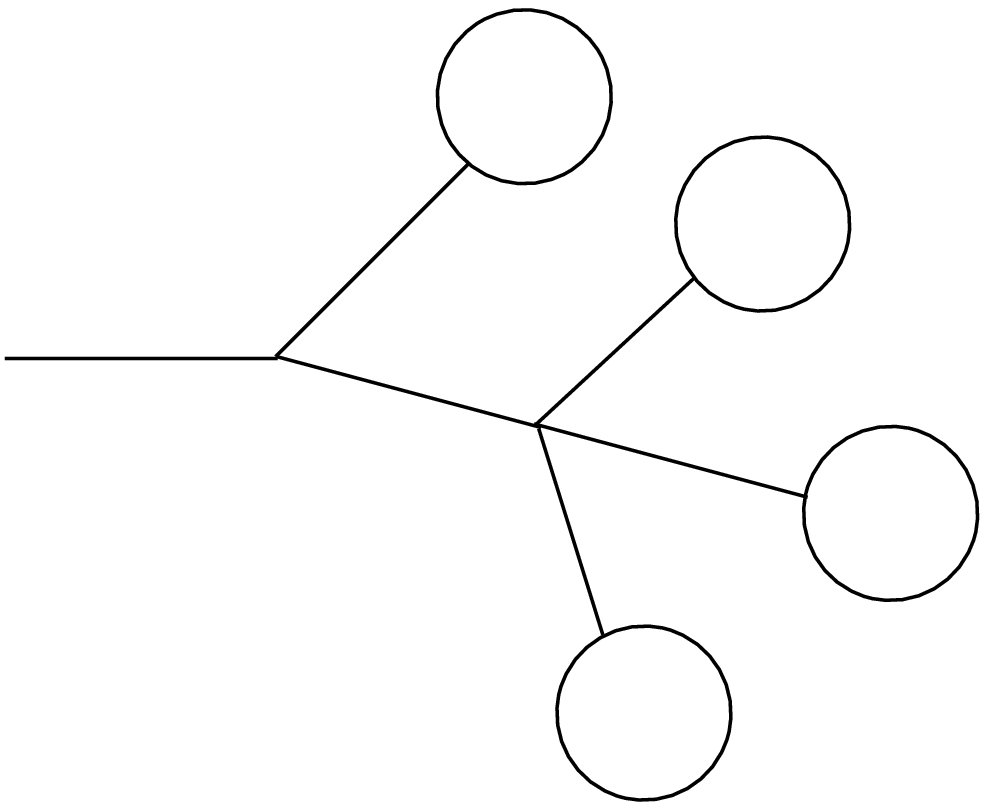}\end{array}
+\cdots .
\end{eqnarray}
Here the blob corresponds to the tadpole $\bar T_0$ and each line
represents the graviton propagator at zero momenta
\begin{eqnarray}\label{eq11}
D_{\mu\nu;\alpha\beta}&=&\frac{1}{2\Lambda}\frac{\eta_{\mu\alpha}\eta_{\nu\beta}
+\eta_{\mu\beta}\eta_{\nu\alpha}-\eta_{\mu\nu}\eta_{\alpha\beta}}{2}
\nonumber \\
&\equiv& \frac{1}{2\Lambda} P_{\mu\nu;\alpha\beta}.
\end{eqnarray}
One can verify, with the help of Eq. \eqref{eq08}, that by summing 
the series of tree diagrams in Eq. \eqref{eq10}, 
one arrives at a result which is equivalent to the one given by Eq. \eqref{eq05}.

The effective potential \eqref{eq02} can have a local extremum at the
value given by Eq. \eqref{eq05}, provided 
$\langle g^{\mu\nu}\rangle$ is a real quantity. This condition
requires that $\det{T_0} < 0$ which entails, together with
Eq. \eqref{eq03}, the relation
\be\label{eq11b}
T < T_{\rm cr} = \left(\frac{45 m^2 \phi_c^2}{\omega \pi^2} \right)^{1/4}.
\ee
This result, which is reminiscent of the one found in $(\phi^3)_6$ theory
\cite{Altherr:1991fu}, indicates the existence of a critical
temperature above which the effective potential \eqref{eq02} has no
extrema. 

Moreover, in order for a local minimum of the effective
potential to exist, it is also necessary that
\be\label{eq13} 
P_{\mu\nu;\alpha\beta}\left.\frac{\delta^2 V^{\rm eff}}{\delta g_{\mu\nu}\delta
g_{\alpha\beta}} \right|_{g^{\mu\nu}=\langle g^{\mu\nu}\rangle} > 0, 
\ee 
where $P_{\mu\nu;\alpha\beta}$ is the projection operator
introduced in Eq. \eqref{eq11}. After a straightforward algebra, one
finds that this requirement leads to the relation 
\be\label{eq12b}
\Lambda\left[\left(\frac{\omega \pi^2 T^4}{15}\right)^2 - 15 m^4
  \phi_c^4\right] > 0.
\ee
In view of \eqref{eq11b}, this inequality implies that $\Lambda$ must
be a negative quantity, in order to ensure the existence of a local
minimum of the effective potential in the tadpole approximation.

We mention here that the constant vacuum expectation value of the
scalar field may be obtained, as discussed for instance in ref. 
\cite{Altherr:1991fu}, through the introduction of self-interactions
between the scalar fields. Such expectation values would then
correspond to the value which minimize the corresponding effective
potential. In this note, for simplicity, we do not consider these
self-interactions, since their inclusion would not alter the essence
of the above considerations. 

\section{Higher order contributions to the effective potential}
In quantum gravity, the leading contributions at high temperature 
satisfy simple Ward identities which reflect the
gauge invariance of the theory
\cite{Frenkel:1991dw,Rebhan:1991yr,Brandt:1992qn}. 
These identities relate  recursively all higher order Green functions
to the one-point (tadpole) function. As a result, any $n$-point Green functions
will exhibit the same leading $T^4$ dependence as does the tadpole
diagram. All these thermal Green functions can be
systematically generated, in the static case, by the effective action
\be\label{eq15}
S_{\rm th} = \frac{\omega\pi^2 T^4}{90}\int {\rm d}^4 x \frac{\sqrt{-g}}{g_{00}^2} 
\ee
which is gauge invariant under static gauge transformations.

For consistency, all such higher order terms must be included in the
effective potential \eqref{eq01}, in which case ${\cal{H}}$ 
may be written for collisionless thermal matter, as
\begin{eqnarray}\label{eq16} 
{\cal{H}} =\frac{m^2 \phi_c^2}{2}\sqrt{-g}
-\frac{\omega\pi^2 T^4}{90}\frac{\sqrt{-g}}{g_{00}^2}.  
\end{eqnarray}
We emphasize that the collisionless approximation may hold at
temperatures such that $G T^2\ll 1$, but would not be justified for temperatures
at the Planck scale, where collisions in the thermal matter become
very important.
Then, the
Euler-Lagrange equation derived from this effective potential may be
written as 
\be\label{eq17} \frac{\Lambda}{8\pi G} g^{\mu\nu} = -
T^{\mu\nu}, 
\ee 
where the complete thermal energy-momentum tensor is given by
\begin{eqnarray}\label{eq18} 
T^{\mu\nu} &=& \frac{2}{\sqrt{-g}}\frac{\delta}{\delta
  g_{\mu\nu}}\int{\rm d}^4 x {\cal{H}}
=\frac{m^2 \phi_c^2}{2} g^{\mu\nu}
\nonumber \\
&+&  \frac{\omega \pi^2 T^4}{90}
\frac{4\delta^\mu_0\delta^\nu_0-g_{00} g^{\mu\nu}}{(g_{00})^3} .  
\end{eqnarray}
We note that the thermal part of the tensor $T^{\mu\nu}$ is a traceless
function at high temperature. Thus, by taking the trace of
Eq. \eqref{eq17}, one would get that 
\be
\Lambda = - 4\pi G m^2 \phi_c^2.
\ee

In view of this condition, we see that Eq. \eqref{eq17} cannot have a 
consistent solution $\langle g^{\mu\nu}\rangle$ at nonzero
temperature. Consequently, we may conclude that the complete thermal effective
action does not have a local minimum.

This behavior may also be seen from a diagrammatic approach similar to the one which led to Eq. \eqref{eq10}. 
In the present case, one finds that the vacuum expectation value of 
the gravitational field can be related to a series of diagrams of the form
\begin{eqnarray}\label{eq20}
 \frac{1}{2} \begin{array}{c}\includegraphics[scale=0.2]{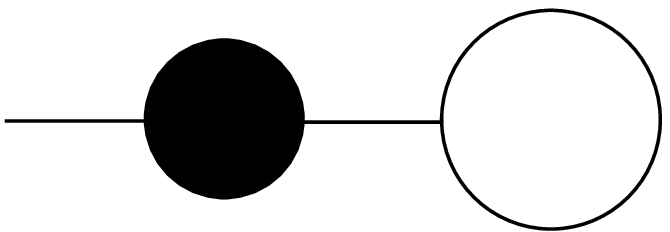}\end{array}
+\frac{1}{8} \begin{array}{c}\includegraphics[scale=0.2]{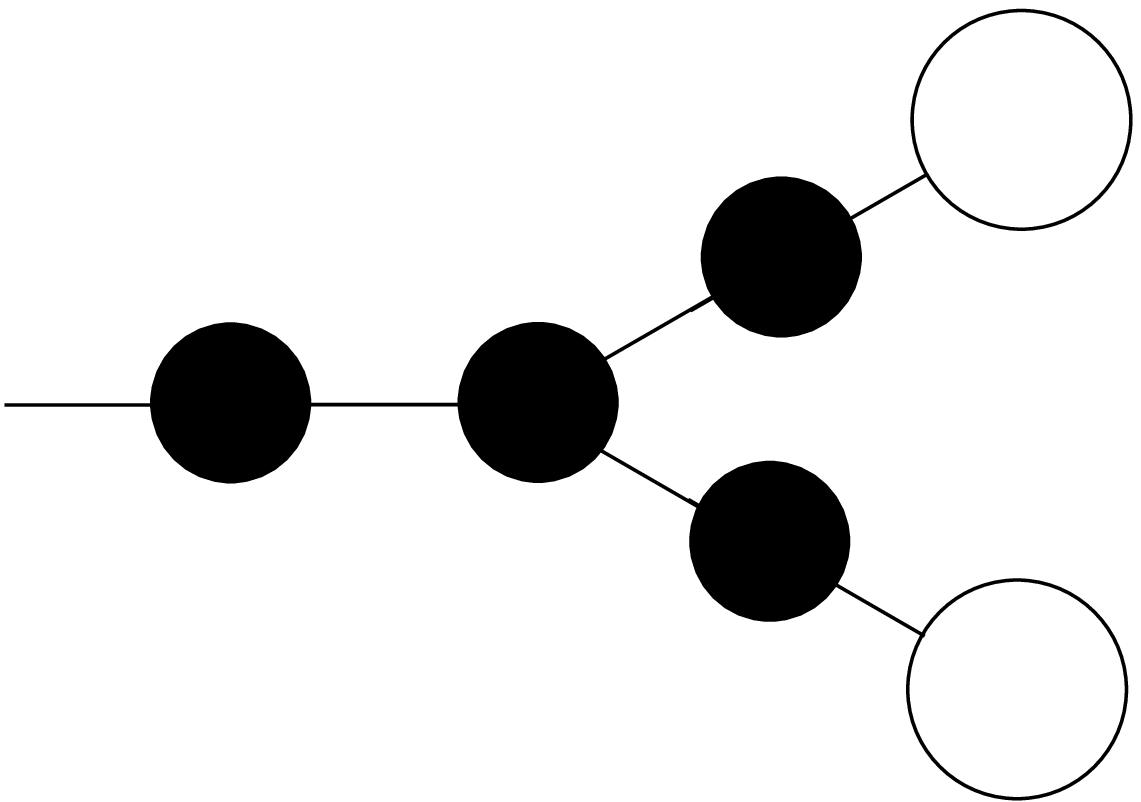}\end{array}
+\cdots
\end{eqnarray}
where the full blobs denote the hard thermal self-energy and vertex
corrections to the tadpole tree-amplitudes. But it may
be checked that the sum of the above series of diagrams would lead 
to a divergent result, so that a regular
solution $\langle g^{\mu\nu}\rangle$ does not exist in this case.

The fact that higher order thermal corrections destabilize the local 
minimum of the effective potential may be understood on physical
grounds. This is because such contributions lead to the appearance 
of an imaginary part in the free-energy, which can be related to the 
decay rate of the quantum vacuum. Work on this aspect is in progress.

\noindent{\bf Acknowledgments}

We would like to thank Professors Ashok Das and J. C. Taylor for
helpful discussions. We are grateful to CNPq and FAPESP, Brazil,
for financial support.

\end{document}